\documentclass[11pt,a4paper,twocolumn]{article}

\usepackage{graphicx}

\usepackage{amssymb}
\usepackage{amsmath}

\usepackage{booktabs}
\usepackage{wasysym}
\usepackage{gensymb}
\usepackage{xcolor}
\usepackage{subfig}

\newcommand\kgd{$\, \rm{kg \cdot d}$ }
\newcommand\kgdmath{\, \rm{kg \cdot d} }
\newcommand\cl[1]{#1\,\%\,C.L.}
\newcommand\frejus{Fr\'{e}jus }
\newcommand\muveto{$\mu$-veto system }
\newcommand\muvet{$\mu$-veto system}
\newcommand\dm{dark matter }
\newcommand\muind{muon-induced }
\newcommand\Muind{Muon-induced }
\newcommand\EDW{EDELWEISS}
\newcommand\EDD{EDELWEISS-II}
\newcommand\EDDD{EDELWEISS-III}
\newcommand\muAcceptance{detection efficiency}

\usepackage{lineno}

\newcommand{\captionfonts}{\footnotesize} 

\makeatletter  
\long\def\@makecaption#1#2{%
  \vskip\abovecaptionskip
  \sbox\@tempboxa{{\captionfonts #1: #2}}%
  \ifdim \wd\@tempboxa >\hsize
    {\captionfonts #1: #2\par}
  \else
    \hbox to\hsize{\hfil\box\@tempboxa\hfil}%
  \fi
  \vskip\belowcaptionskip}
\makeatother   

\long\def\symbolfootnote[#1]#2{\begingroup%
\def\thefootnote{\fnsymbol{footnote}}\footnote[#1]{#2}\endgroup}

\begin{document}

\onecolumn
\begin{center}
\LARGE 
Muon-induced background in the \EDW\ \dm search
\end{center}

\vspace{5mm}

\begin{center}
\small
The EDELWEISS collaboration,\\
B.~Schmidt\symbolfootnote[1]{Corresponding author, b.schmidt@kit.edu}$^{,\;a)}$,
E.~Armengaud$\;^{b)}$,
C.~Augier$\;^{c)}$,
A.~Benoit$\;^{d)}$,
L.~Berg\'e$\;^{e)}$,
T.~Bergmann$\;^{f)}$,
J.~Bl\"umer$\;^{a),\;g)}$,
G.~Bres$\;^{d)}$,
A.~Broniatowski$\;^{e)}$,
V.~Brudanin$\;^{h)}$,
B.~Censier$\;^{c)}$, 
M.~Chapellier$\;^{e)}$,
F.~Charlieux$\;^{c)}$,
S.~Collin$\;^{e)}$,
P.~Coulter$\;^{i)}$,
G.A.~Cox$\;^{a)}$,
O.~Crauste$\;^{e)}$,
J.~Domange$\;^{b),\;e)}$,
L.~Dumoulin$\;^{e)}$,
K.~Eitel$\;^{g)}$, 
D.~Filosofov$\;^{h)}$,
N.~Fourches$\;^{b)}$,
G.~Garde$\;^{d)}$,
J.~Gascon$\;^{c)}$,
G.~Gerbier$\;^{b)}$,
M.~Gros$\;^{b)}$,
L.~Hehn$\;^{g)}$,
S.~Henry$\;^{l)}$,
S.~Herv\'e$\;^{b)}$,
G.~Heuermann$\;^{a)}$,
A.~Juillard$\;^{c)}$,
H. Kluck$\;^{a)}$, 
V.Y.~Kozlov$\;^{g)}$,
M.~Kleifges$\;^{f)}$,
H.~Kraus$\;^{i)}$,
V.A.~Kudryavtsev$\;^{j)}$,
P.~Loaiza$\;^{k)}$,
S.~Marnieros$\;^{e)}$, 
A.~Menshikov$\;^{f)}$,
X.-F.~Navick$\;^{b)}$,
H.~Nieder\symbolfootnote[2]{current address: Karlsruhe Institute of Technology, Institut f\"ur Meteorologie und Klimaforschung, 76021 Karlsruhe, Germany}$\;^{a)}$,
C.~Nones$\;^{b)}$,
E.~Olivieri$\;^{e)}$,
P.~Pari$\;^{l)}$,
B.~Paul$\;^{b)}$,
M.~Robinson$\;^{j)}$,
H.~Rodenas$\;^{d)}$,
S.~Rozov$\;^{h)}$,
V.~Sanglard$\;^{c)}$,
B.~Siebenborn$\;^{g)}$,
D.~Tcherniakhovski$\;^{f)}$,
A.S.~Torrent\'o-Coello$\;^{b)}$,
L.~Vagneron$\;^{c)}$,
R.J.~Walker$\;^{a)}$,
M.~Weber$\;^{f)}$,
E.~Yakushev$\;^{g)}$,
X.~Zhang$\;^{i)}$
\end{center}

\vspace{4pt}

\begin{center}
\footnotesize \it 
\noindent $^{a)}${Karlsruhe Institute of Technology, Institut f\"ur Experimentelle Kernphysik, Gaedestr. 1, 76128 Karlsruhe, Germany}\\
\noindent $^{b)}${CEA, Centre d'\'Etudes Nucl\'eaires de Saclay, IRFU, 91191 Gif-sur-Yvette Cedex, France}\\
\noindent $^{c)}${Institut de Physique Nucl\'eaire de Lyon, Universit\'e de Lyon (Universit\'e Claude Bernard Lyon 1) et IN2P3-CNRS, 4 rue Enrico Fermi, 69622 Villeurbanne, France}\\
\noindent $^{d)}${Institut N\'eel, CNRS, 25 Avenue des Martyrs, 38042 Grenoble Cedex 9, France}\\
\noindent $^{e)}${Centre de Spectroscopie Nucl\'eaire et de Spectroscopie de Masse, UMR8609 IN2P3-CNRS, Univ. Paris Sud, b\^at 108, 91405 Orsay Campus, France}\\
\noindent $^{f)}${Karlsruhe Institute of Technology, Institut f\"{u}r Prozessdatenverarbeitung und Elektronik, Postfach 3640, 76021 Karlsruhe, Germany}\\
\noindent $^{g)}${Karlsruhe Institute of Technology, Institut f\"ur Kernphysik, Postfach 3640, 76021 Karlsruhe, Germany}\\
\noindent $^{h)}${Laboratory of Nuclear Problems, JINR, Joliot-Curie 6, 141980 Dubna, Moscow Region, Russian Federation}\\
\noindent $^{i)}${University of Oxford, Department of Physics, Keble Road, Oxford Ox1 3RH, UK}\\
\noindent $^{k)}${University of Sheffield, Department of Physics and Astronomy, Sheffield S3 7RH, UK}\\
\noindent $^{k)}${Laboratoire Souterrain de Modane, CEA-CNRS, 1125 route de Bardonn\`eche, 73500 Modane, France}\\
\noindent $^{l)}${CEA, Centre d'\'Etudes Nucl\'eaires de Saclay, IRAMIS, 91191 Gif-sur-Yvette Cedex, France}\\

\end{center}

\vspace{4pt}

\begin{abstract}
\small
A dedicated analysis of the \muind background in the \EDW\ \dm search has been performed on a data set acquired in 2009 and 2010. The total muon flux underground in the Laboratoire Souterrain de Modane (LSM) was measured to be $\Phi_{\mu}=(5.4\pm 0.2 ^{+0.5}_{-0.9})$\,muons/m$^2$/d. The modular design of the \muveto allows the reconstruction of the muon trajectory and hence the determination of the angular dependent muon flux in LSM. The results are in good agreement with both MC simulations and earlier measurements. 
Synchronization of the \muveto with the phonon and ionization signals of the Ge detector array allowed identification of \muind events. Rates for all \muind  events  $\Gamma^{\mu}=(0.172 \pm 0.012)\, \rm{evts}/(\rm{kg \cdot d})$ and of WIMP-like events $\Gamma^{\mu-n} = 0.008^{+0.005}_{-0.004}\, \rm{evts}/(\rm{kg \cdot d})$ were extracted. After vetoing, the remaining rate of accepted \muind neutrons in the \EDD\ \dm search was determined to be $\Gamma^{\mu-n}_{\rm irred} < 6\cdot 10^{-4} \, \rm{evts}/(\rm{kg \cdot d})$ at 90\%\,C.L. Based on these results, the \muind background expectation for an anticipated exposure of 3000\,\kgd\ for \EDDD\ is $N^{\mu-n}_{3000 kg\cdot d} < 0.6$ events. 
\end{abstract}

\begin{flushleft}
\end{flushleft}

\twocolumn
\section{Introduction}
\label{sec:Intro}
Astrophysical and cosmological observations \cite{Pe96, Kom11} have led to the current cosmological concordance model that incorporates both dark energy and \dm as driving forces in the evolution of the Universe. Dark matter herein constitutes a matter-like fluid ruling the dynamics of structure formation as well as significantly affecting the dynamics of our own galaxy today \cite{Ka98}. Assuming extensions of the Standard Model with a new particle originally in thermodynamic equilibrium, but decoupling shortly after the Big Bang, a promising class of \dm candidates called WIMPs (Weakly Interacting Massive Particles) arises. An annihilation cross section of the order of the weak scale is needed to produce the observed cosmic abundance \cite{Ber05}, rather independent of the WIMP mass which is typically assumed to be of O(100\,GeV). With these properties, this candidate is non-relativistic at freeze-out and hence consistent with our understanding of structure formation.

The challenge of direct \dm search experiments is the detection of the extremely rare WIMP-nucleus scattering events. Furthermore the expected energy spectrum from WIMP-nucleus scattering exhibits an exponential distribution, tailing off towards higher energy and can be easily mimicked by background sources. Thus, in order to ensure that a signal is caused by WIMPs one has to understand all background sources and to suppress as many as possible. The overwhelming majority of observed events are gamma and beta particles from ambient radioactivity. These are rejected with the powerful discrimination capabilities offered by the simultaneous readout of two channels such as phonons and scintillation \cite{Angloher2009,Ang11,Brown2012}, or phonons and ionization \cite{Ahmed2011,Ahm11}, or scintillation and ionization \cite{Plante2011,Aprile2012,Akimov2011}, and fiducialization (bulk - surface) in some detectors. In contrast, a neutron background causes a WIMP-like signal (a nuclear recoil) in the detectors and can 
only 
be discriminated via multiple scattering of neutrons or coincident $\mu$-veto signals if induced by cosmic muons.

At levels of sensitivity to spin-independent elastic WIMP-nucleon cross sections of a few $10^{-44}$\,cm$^2$ (CDMS and \EDW\  with 614\,\kgd of exposure \cite{Ahm11}) or lower (XENON100 \cite{Aprile2012} with 2323.7\,\kgd equivalent exposure), further reducing the background is a demanding task. It requires a careful analysis of all known backgrounds and the extrapolation to the next stages of experiments \cite{EDWbg12}. For \EDW, prominent sources of backgrounds are a potential failure to separate electron recoils from nuclear recoils via the charge and phonon measurement, and neutrons produced via ($\alpha$, n) reactions, spallation as well as neutrons induced by muons. 
Old results have considerable systematic uncertainties due to the lack of detailed Monte Carlo modeling of muon transport, secondary particle production and detection within a specific detector configuration. More recent experiments (see for instance \cite{AbeKamland10,Boulby08,Borexino11,Mar07}) provide more reliable data for different targets but some of them are still not consistent with the existing models based on the GEANT4 toolkit \cite{GEANT4-03} or FLUKA code \cite{Batti06} (see also simulations and discussions in Refs. \cite{Ara05, Lindote08}).

In this article we present the results of an investigation of the \muind background in the \EDW\ experiment. We start with an introduction into the experimental setup, describing briefly the Ge bolometers and focussing on the \muveto and its data acquisition (Section~\ref{sec:Setup}). In Section~\ref{sec:DetectMuons}, we derive the muon detection efficiency and reconstruct muon information such as angular distributions and track coordinates. Based on these data, a search for coincidences of muon veto and bolometer events has been performed with results presented in Section~\ref{sec:Coincs}. We conclude with an outlook for the \muind background expectation for the upcoming \EDDD\ \dm search.

\section{Experimental setup}
\label{sec:Setup}

\begin{figure*}[htb]
 \begin{center}
 \includegraphics[width=0.9\textwidth]{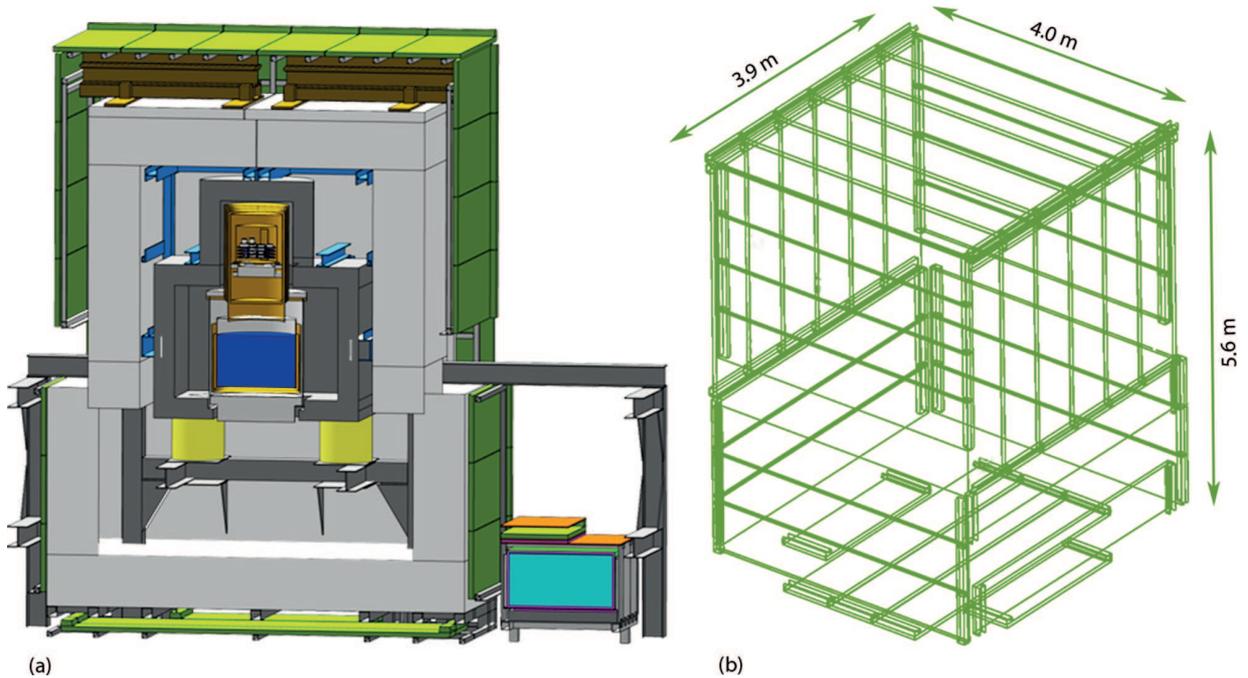}
\end{center}
 \caption{(a) Cross-section of the \EDW\ experimental setup as implemented in the MC simulations. (b) Schematic view of the modular \muvet.}
 \label{fig:Setup}
\end{figure*}

To shield against cosmic backgrounds, the \EDW\ experiment is housed in the Laboratoire Souterrain de Modane (LSM), an underground laboratory under the French Alps with an average rock overburden of 4800\,m w. e. (meters water equivalent). This attenuates the muon flux by six orders of magnitude compared to sea level down to $\sim5\,\mu/\rm{m^2/d}$. The experimental setup depicted in Fig.~\ref{fig:Setup} consists of several layers of active and passive shielding surrounding the innermost part, a dilution cryostat. This cryostat with a working temperature of 18\,mK contains the ultrapure Ge crystals used as detectors for the WIMP search.

The outermost shielding is an active \muveto of plastic scintillator modules, described in Section~\ref{sec:VetoSystem}. It is mounted on a 50\,cm thick polyethylene (PE) layer, inside which the actual cryostat housing is shielded with 20\,cm thick lead, of which the inner 2\,cm are ancient lead (Pb). The PE moderates the neutron energies to sufficiently low levels for a significant fraction of them to be absorbed. The innermost 2\,cm of the lead were salvaged from the wreck of an ancient Roman galley \cite{Ho87} in order to reduce the 46.5\,keV $\gamma$-line of the $^{210}$Pb as well as X-rays and bremsstrahlung photons from the $\beta^-$ decay of $^{210}$Bi. The Ge bolometers are described in Section~\ref{sec:BolometerSystem}. Their signals are sampled using a common clock with a 100\,kHz frequency. That same clock is transferred to the \muveto data acquisition (DAQ) via optical fibers. Apart from this synchronization, both systems run individually and muon-induced bolometer events are identified offline
within the main analysis \cite{Arm11}. An additional 1~ton liquid scintillator detector designed to study muon-induced neutrons with better statistics was installed at the end of 2008 (Fig.~\ref{fig:Setup}a, detector outside the shielding at the lower right). The detailed setup and first results obtained with this additional detector are described in \cite{Koz10}. Further measurements of the ambient neutron flux are performed inside and outside of the shielding with bare $^3$He neutron counters \cite{Roz10,Eit12}.

\subsection{Bolometer array and DAQ}
\label{sec:BolometerSystem}
The Ge crystals for \EDD\ are grown from ultrapure Ge with less than 10$^{10}$ impurities per cm$^3$ and cut as cylinders of 70\,mm diameter and 20\,mm height. The simultaneous readout of a heat signal with a neutron transmutation doped (NTD) Ge sensor \cite{Mar04, Mar08}, and the ionization with charge collection electrodes enables the distinction between electron and nuclear recoils. During the data taking period 2009-2010, the dilution cryostat provided a stable detector temperature of 18\,mK for more than one year. Five ID detectors with beveled edges and a mass of 360\,g each were used together with five plain cylindrical detectors of 410\,g mass. These ID detectors were equipped with interdigitized collecting and veto electrodes \cite{Bro09}. The Al electrodes are evaporated on the top and bottom of the Ge crystal. Each ring has a width of 200\,$\mu$m and 1.8\,mm pitch, and is typically biased at voltages of $\pm4$\,V for the collecting and $\mp1.5$\,V for the veto electrode. Solid guard electrodes
covering the outer rim and the beveled faces allow suppression of events in the outer regions of the crystals. The electrical field lines run orthogonal to the surface for bulk events, but parallel for surface events. This design therefore allows to exclude events in the surface layers of the detector, which are prone to charge collection deficiencies.

In the analysis the discrimination between electron and nuclear recoils is based on the ionization yield Q, which is the ionization energy (calibrated for electron recoils with the 356\,keV line of $^{133}$Ba) relative to the recoil energy. For electron recoils the value of Q is set to unity by calibration. Nuclear recoils transmit most of their energy directly to vibrations of the crystal lattice (phonons), without going through ionization processes. Thus, they generally have a much smaller ionization yield than electron recoils. For the \EDW\ detectors, the ionization yield of nuclear recoils as a function of energy follows closely the prediction of the Lindhard theory \cite{Lin63} and can be parameterized through the expression, determined in an extensive neutron calibration \cite{Mar04,Ste01}:
\begin{linenomath}
\begin{align}
 Q_{\rm{Lin}}=0.16\cdot (E_{\rm{Rec}}[\rm{keV}])^{0.18}.
\end{align}
\end{linenomath}
 Assuming a Gaussian uncertainty of the $Q$ value distribution with energy-, detector- and time-dependent resolutions $\sigma_{\rm ion}(\rm{E_{\rm ion},t,det})$ and $\sigma_{\rm heat}(\rm{E_{\rm heat},t,det})$, a recoil band can be defined based on the standard deviation
\begin{equation}
  \sigma_Q = \sigma_Q(E_{\rm{Rec}},\sigma_{\rm ion}(\rm{t,det}),\sigma_{\rm heat}(\rm{t,det})).
\end{equation}
For the ID detectors, rejection powers of better than 1 in 10$^5$ for $\gamma$ rays in the bulk and 1 in 6$\cdot 10^4$ for surface events were determined using a $^{133}$Ba $\gamma$ and a $^{210}$Pb $\beta$ source, respectively \cite{Bro09}. Whilst these rejection powers are very good, the surface event rejection together with the rejection of events on the radial surface, so-called guard events, significantly reduces the volume of accepted events. The bulk or fiducial volume, where reliable charge collection is expected, was defined in the following way. Each of the signals on veto and guard electrodes, as well as the charge imbalance between the two collecting electrodes had to be consistent with zero at 99\,\%C.L., according to the resolution of these parameters. The efficiency of those cuts was measured using the low-energy 8.98\,keV and 10.34\,keV peaks associated with the cosmogenic activation of germanium, producing a uniform contamination with the isotopes $^{65}$Zn and $^{68}$Ge. Averaged over time
and all detectors, a typical fiducial mass of $(166\pm6)\,$g per detector was found \cite{Arm11}.

With a live data taking period of 307 days during the cryogenic run from March 2009 to May 2010, an effective fiducial exposure of $(481\pm 17)$ \kgd has been collected for the analysis of the \muind background in the bolometers including a factor of 0.9 to account for the region of interest (RoI) being the \cl{90} nuclear recoil band. This exposure is larger than the 384 \kgd used in the WIMP search in \cite{Arm11}, as the present study of coincidences of muon veto and germanium detector events does not require the same strict quality cuts on the noise levels in the germanium detectors. While the fiducial volume cut is paramount for the DM search \cite{Arm11} it can be beneficial to include less well reconstructed events closer to the surface of the detectors to increase statistics for coincidence studies. Therefore, we also present the results of a coincidence study using looser cuts for events in the whole volume of the detectors, for which the exposure amounts to 1504\,kg$\cdot$d.

\subsection{Muon veto system}
\label{sec:VetoSystem}
The outermost layer of the \EDW\ setup is the $\mu$-veto system, a layer of plastic scintillator acting as an active shield to tag any muon-induced activity in the bolometer system. It consists of 46 individual plastic scintillator modules (Bicron BC-412) of 65\,cm width, 5\,cm thickness and lengths of 2\,m, 3.15\,m, 3.75\,m and 4\,m adding up to a total surface of 100\,m$^2$. The modules are assembled in a stainless steel mounting frame and attached to the PE shielding to ensure an almost hermetic coverage of the experiment setup. Due to the need for access to the cryostat during maintenance intervals, parts of the \muveto are positioned on rails on top of the two movable PE blocks to allow a roll-back during cryostat access (Fig.~\ref{fig:Setup}). Whereas the upper level (N1) of the \muveto almost completely covers the PE shielding, except for small openings for rails, the part below the cryostat level (N0) has more prominent gaps due to the cryogenic supply lines, and the pillars on which the entire
experiment is mounted. The overall geometrical muon flux coverage of the \muveto amounts to 98\,\%. To ensure proper closure during data taking, the exact positions of the two mobile veto parts are monitored via regular position measurements using laser interferometry.

Each module is viewed by two groups of four 2-inch PMTs (Philips Valvo XP2262/PA). Each group has an individual high voltage (HV) supply and readout, leading to a total of 92~HV and signal channels. Similar modules have been used as a muon veto counter in the KARMEN experiment \cite{KARMEN02}. A complete description of the modules including measurements of the effective attenuation length and spectral quantum efficiency of the scintillator can be found in \cite{Rei98}. The data acquisition is based on VME electronics with special cards developed in-house \cite{Hab04}. An event recording in the veto system is triggered once the two signals of a module each pass a trigger threshold within a coincidence window of 100\,ns. With leading edge discriminators set at the same level of 150\,mV, individual modular thresholds are achieved by adjusting the HV values. Within an event, all hits above threshold in the modules are then registered within an interval of $\pm125$\,ns around the trigger time. For each signal
channel, the individual time of a hit is recorded in a 128-channel TDC (CAEN V767) and the analog signals are converted in three 32-channel QDCs (CAEN V792). After the coincidence window, any further input is blocked for about 50\,$\mu$s \cite{Hab04} until the event data are read out via the VME bus and all active elements are reset by the acquisition software. The time information with respect to the bolometers is received via an optical fiber which transmits the 10\,$\mu$s time stamp in 48 bits of 64\,ns width each. This sequence of bits is converted via an FPGA-based board and then read out via the VME bus. To monitor the overall status of the veto modules, the hit rate of all PMT groups (not requiring any coincidence) is recorded in scalers. These data are processed continuously to check the performance of the \muveto via a graphical web interface.

\begin{figure}[tb]
 \centering
	\includegraphics[width=0.48 \textwidth]{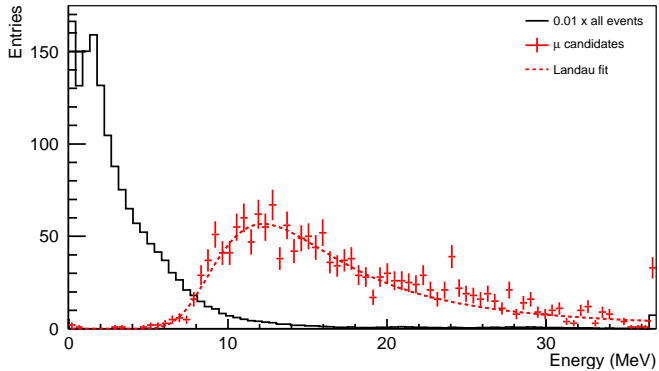}
 \caption{Distribution of energy depositions in a top module. The histogram (solid black) shows events recorded for a standard trigger condition (internal coincidence of the module), scaled by 0.01. Data points (in red) describe energy depositions of muon candidates selected by requiring full non-zero ADC and TDC information in another module (two modules in coincidence) and follow a Landau distribution.}
 \label{fig:MuBackgroundEnergy}
\end{figure}

A muon traversing a scintillator module as a minimum ionizing particle deposits on average more than 10\,MeV (see Fig.~\ref{fig:MuBackgroundEnergy} for a typical Landau spectrum). The average energy deposition is 11.8\,MeV for horizontal modules on the top and bottom, and up to 24\,MeV for modules on the sides \cite{Ho07}. As seen in Fig.~\ref{fig:MuBackgroundEnergy}, these values are above typical energy deposition from the ambient radioactivity. For each channel, the energy deposition is stored as the integrated charge Q in units of ADC values. With the comparison of the Landau most probable values (MPV) of the integrated charge Q of specific two module coincidence cuts with dedicated MC simulations, a module-individual calibration is obtained with a mean calibration coefficient of 5.44\,keV/ADC-value. The effective threshold of the muon veto modules has been chosen to be much lower than the MPVs (about 5\,MeV at the center of a module, rising towards the module ends due to the position-dependent 
light
output) to be also able to detect grazing muons and thus to keep the muon detection efficiency as high as possible. As a result, the rate of recorded events in the entire system is of the order of one per second, to be compared to an expected rate of throughgoing muons hitting at least one module per system section N1 and N0 of $\sim 3.5\times 10^{-4}$\,Hz. Whereas the rate of recorded events is dominated by ambient background, candidates of throughgoing muons can be easily extracted from the data by requiring two modules in coincidence.

Vetoing a bolometer event due to activity in the \muveto is performed in the offline analysis. A veto is applied in a conservative time window of $\pm 1$\,ms around each event recorded in the $\mu$-veto system, regardless of the fact that these events are dominated by ambient background. This offline reduction of effective exposure due to the entire \muveto results in a deadtime of the order of 0.2\,\%. As the veto system is completely independent of the bolometer system and its data acquisition, data of these scintillator modules are acquired continuously, only interrupted by maintenance intervals.

\begin{figure}[tb]
 \centering
	\includegraphics[width=0.48 \textwidth]{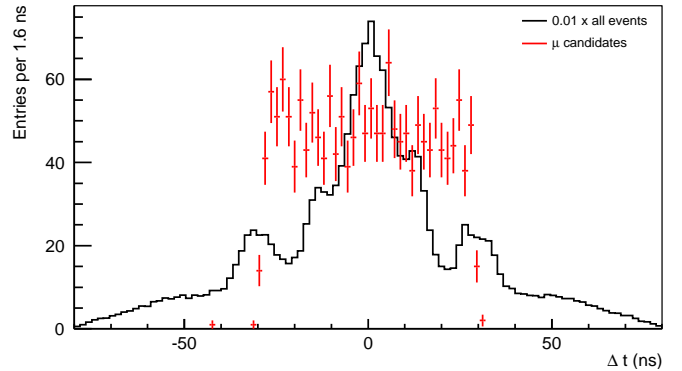}
 \caption{Time difference of PMT signals within a top module for events with standard trigger condition (internal coincidence) in solid black, scaled by 0.01, and for muon candidates (red data points), selected with a two module coincidence cut. }
 \label{fig:MuBackgroundTDC}
\end{figure}
For each event in the $\mu$-veto system, the position of an energy deposition can be reconstructed using the module number and the time difference between the signal arrival at each module end. The time is stored in TDC units, where one TDC unit corresponds to 0.8\,ns. A typical distribution of the time difference within a $\mu$-veto module is shown in Fig.~\ref{fig:MuBackgroundTDC}. The black distribution recorded with the standard trigger condition shows the threshold dependence along the module length with the minimum threshold and largest count rate at the center of the module. Additional contributions are visible at $\Delta t \approx \pm 30$\,ns due to a thicker scintillator and thus higher detection efficiency at the module ends. The tails to much larger time differences are due to the limited time resolution for low energy deposition as well as to light emission in the PMTs. In contrast, muon candidates show a much more confined distribution of time differences as can be seen in Fig.~\ref{fig:MuBackgroundTDC}. The rather flat distribution with sharp edges reflects the physical length of a module with homogenously distributed hits along the module axis. Thus, with the event data on energy deposition and time, sufficient information is provided to perform cuts selecting muon candidates and to reconstruct muon trajectories (see Section~\ref{mtrackreconst}).

\section{Detecting muons}
\label{sec:DetectMuons}

In this section, we derive the detection efficiency of the \muveto in two different ways. Based on a detailed simulation of muons and their interactions in the \EDW\ setup (Section~\ref{sub:Geant4Setup}) using individual modular efficiency curves determined in situ from data (Section~\ref{sub:DetectionEfficiency}), the first method is a combination of MC results and data (Section~\ref{sub:resultMC}). The second method uses bolometer data identified as \muind events searching for a coincident signal in the \muveto (Section~\ref{sub:Eff-from-data}) and thus is entirely based on data extracted during the WIMP search itself. The measured angular distribution of muons (Section~\ref{mtrackreconst}) as well as the total muon flux are determined (Section~\ref{sec:muonFlux}).

\subsection{Simulation of muons in LSM}
\label{sub:Geant4Setup}
\begin{figure}[tb]
 \centering
 \includegraphics[width=0.48 \textwidth, trim= 2mm 0mm 2mm 0mm]{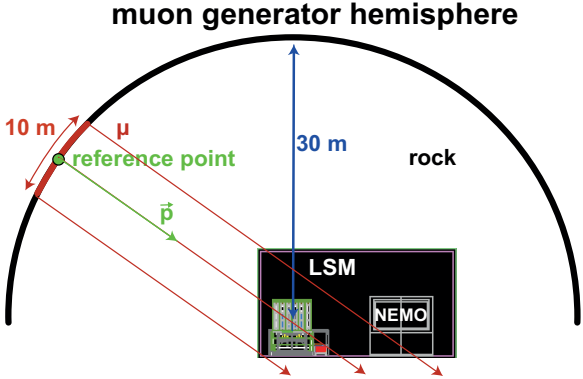}
 \caption{LSM geometry and muon generation as implemented in Geant4. The hemisphere for muon generation is placed in the rock surrounding the laboratory hall with the \EDW\ and NEMO setups. The NEMO-III experiment is implemented as simplified PE shields whereas the \EDD\ setup is implemented with all details (see also Fig.~\ref{fig:Setup}).}
 \label{fig:SimSetup}
\end{figure}

Modeling of muons and their interactions is performed with the Geant4 package version 4.9.2 \cite{GEANT4-03, GEANT4-06}. For these simulations, the detailed geometry of the \EDW\ setup as given in Fig.~\ref{fig:Setup} has been extended to incorporate nearby structures in the LSM laboratory, namely the rock surrounding the laboratory and the adjacent NEMO-III experiment. Figure~\ref{fig:SimSetup} shows the implemented geometry as well as the hemisphere at a distance of 30\,m from the center of the experiment on which muons are generated.
The full MC simulation consists of $2\cdot 10^6$ generated muons in the energy range from 2~GeV-200~TeV. 40.9\% of the started muons have a momentum vector that geometrically would hit the \muvet, the rest illuminates the close surroundings. However, only 33.9\% of the muon tracks actually enter the veto volume. The difference of 7\% is due to muons stopped within the rock overburden.

The energy and angular dependent muon flux above the laboratory is generated according to the energy dependent muon spectrum at sea level \cite{Gai90}:
\begin{align}
 \frac{dN_0}{dE_0} = &\frac{0.14(E_{\mu}/\rm{GeV})^{-\gamma}}{\rm{cm^2~s~sr~GeV}} \cdot   \nonumber \\
&\left( \frac{1}{1+\frac{1.1E_{\mu} cos\Theta}{115~\rm{GeV}}}+\frac{0.054}{1+\frac{1.1E_{\mu} cos\Theta}{850~\rm{GeV}}}\right),
\end{align}
with the cosmic ray exponent $\gamma=2.7$, which is then modified due to energy loss of muons and the shielding in the rock overburden of LSM. This rock overburden $h(\Theta, \Phi)$ is taken from a detailed elevation map with $1\degree \times 1\degree$ resolution \cite{Wei93}.

The muons are generated according to the resulting angular dependent muon flux on a half sphere of 30~m radius centered around the cryostat of the \EDD\ experiment. In order to ensure a homogeneous illumination of the \muvet, the starting point of a muon with momentum vector $\vec{p}$ is randomized within a distance $\delta \leq 5~m$ of a reference point on the hemisphere.  After generation, particles are propagated through the entire setup based on the QGSP\_BIC\_HP reference physics list of Geant4 version 4.9.2 with an adaptation suggested by \cite{Ho07}. This includes the use of the \textquotedblleft low energy extension\textquotedblright\ packages instead of the standard ones for electron, positron and gamma interactions. 
Further, G4MuNuclear was added to model muon-hadronic interactions, and a CHiral Invariant Phase Space (CHIPS) based model was used for gamma-nuclear physics.

To model hadronic interactions, especially inelastic neutron scattering, two further changes were applied. 
At energies between 20\,MeV and 70\,MeV, the package \textit{G4PreCompoundModel} (PC) mediates the transition from \textit{G4NeutronHPInelastic} (HP, up to 20\,MeV) to \textit{G4BinaryCascade} (BIC, from 70\,MeV to 6\,GeV). To bridge a model gap at energies 6\,GeV~$\leq E_{\rm n,p} \leq 12$\,GeV  between BIC and the high energy Quark Gluon String Precompound (QGSP) model implemented in \textit{G4QGSModel}, the Low Energy Parameterized (LEP) model \textit{G4LENeutronInelastic} was inserted. This description of the physical processes has been proven to be effective in terms of CPU consumption and to be accurate when comparing with experimental results \cite{Ho07, Klu12}. Using these interaction models, the distance of 30~m of rock between starting point and experimental setup allows the development of hadronic and electromagnetic showers into equilibrium and ensures the proper abundance of secondary particles in the simulation. Energy depositions by the primary muon and by all secondary particles are then stored for each individual veto module as well as for each bolometer in the cryostat. To avoid boundary effects, two additional meters of rock below the entire setup are included in the geometry.

\subsection{Trigger efficiency of individual veto modules}
\label{sub:DetectionEfficiency}

To determine the detection efficiency for a given simulated energy deposition in a module, the individual trigger threshold of all modules must be known. Although this is a position dependent function along the axis for a given module $i$, it can be measured in situ as an efficiency $\epsilon_i(E)$ averaging over all hit positions. Such a measurement provides an accurate efficiency curve accounting for the local environment of the underground laboratory, ageing of detector components over the long term measurement as well as adjustments of the high voltage applied to the PMTs.

The measurement of $\epsilon_i(E)$ makes use of the fact that, having an internal coincidence in an individual module (TDC data from both ends within a 100\,ns window), all non-zero ADC and TDC data of other modules are also read out and stored. This additional information allows to deduce the trigger efficiency in the following way:
\begin{figure}[tb]
 \centering
 \includegraphics[width=0.48 \textwidth]{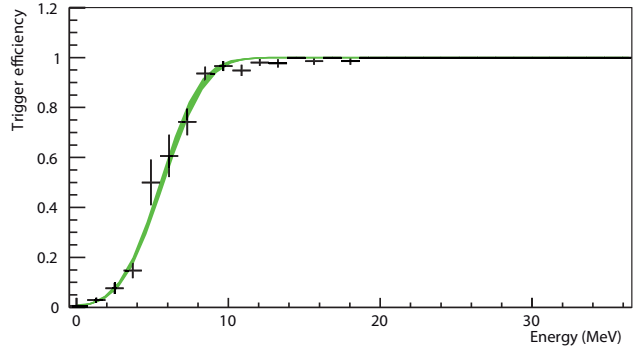}
 \caption{Example of the trigger efficiency $\epsilon_i(E)$ as a function of the energy deposition $E$ in MeV in a module $i$ (shown for a top module, with binomial error estimates). The green band is the 1$\sigma$ region of the efficiency fit, fixing the efficiency level for large energies to $\epsilon_i(E)=1$. The functional form is given by the convolution of a threshold step-function with a gaussian energy resolution.}
 \label{fig:Einzeleffizienz2}
\end{figure}
We select all events where there is an energy deposition in a certain module recorded as non-zero ADC value(s), but where the trigger condition was fulfilled by another module. Whenever both TDC channels of the module under consideration are non-zero, this module would have triggered as well. Otherwise, despite some energy deposition recorded in the ADC, the signals would not have triggered the event recording. The fraction of potential triggers over all events as a function of the calibrated energy is shown in Fig.~\ref{fig:Einzeleffizienz2} as an example of the trigger efficiency $\epsilon_i(E)$ of a specific module $i$.

\begin{figure}[tb]
 \begin{center}
 \includegraphics[width=0.48\textwidth]{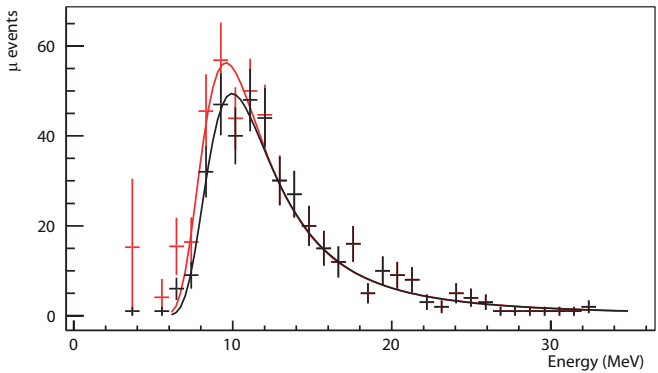}
\end{center}
 \caption{Measured energy deposition $N_{\mu,i}(E)$ of muon candidates for a top module requiring a second module with TDC coincidence (black); Energy spectrum $N^c_{\mu,i}(E)$ after correcting for inefficiencies (red). Both distributions have been fit with a Landau distribution excluding events at low energy which are due to energy depositions of secondary particles.}
 \label{fig:Einzeleffizienz1}
\end{figure}

For muons, one can deduce a detection efficiency $\epsilon^{\mu}_i$ by weighting the energy dependent trigger efficiency with the measured Landau spectrum of each module:
\begin{linenomath}
\begin{align}
    \epsilon^{\mu}_i=\frac{\int_{6 \rm{MeV}}^{\infty} \epsilon_i(E)N_{\mu,i}(E)dE}{\int_{6 \rm{MeV}}^{\infty} N_{\mu,i}(E)dE}.
 \label{eq:modeff}
\end{align}
\end{linenomath}
At energies below 6\,MeV, the selection criterion for a muon candidate, i.e. two non-adjacent modules in coincidence, may also select low-energy secondaries in one of the modules instead of throughgoing muons. The energy cut of 6\,MeV was chosen since the Landau spectra have a negligible fraction of events below that value in contrast to the data with a significant part due to secondaries. Fig.~\ref{fig:Einzeleffizienz1} shows the Landau spectra before and after correction of the efficiency, $N_{\mu,i}(E)$ and $N^c_{\mu,i}(E)=N_{\mu,i}(E)/\epsilon_i(E)$, respectively.

The accuracy of the individual muon detection efficiency is dominated by the uncertainty of the energy calibration based on fitting Landau spectra. The expected Landau MPV varies for different modules due to the mountain profile, the orientation of the modules and the muon selection cut and amounts to $\langle E_{\rm{MPV}}\rangle =16$\,MeV. The Landau distribution itself is significantly broadened in comparison to that of vertically throughgoing muons. The energy calibration per module is based on a modular comparison of MPV's from data fits (in ADC units) with MC simulations in MeV and is estimated to be accurate at the 20\% level. The resulting distribution of energies where the individual module trigger efficiency reaches 50\% is Gaussian with $\langle E(\epsilon_i=0.5)\rangle =7$\,MeV as mean and a standard deviation of $\sigma(E)=1.9$\,MeV. Averaging the modular detection efficiency $\epsilon^{\mu}_i$ in Eq.~\ref{eq:modeff} results in $\langle \epsilon^{\mu}_i \rangle = 95\%$ with $\sigma(\epsilon^{\mu}_
i)=4$\%.

\subsection{Muon \muAcceptance\ derived from MC simulations}
\label{sub:resultMC}

In order to evaluate the muon \muAcceptance\ for the entire \muveto we defined two reference volumes for entering muons. At first, we take the volume defined by the \muveto itself. A simulated energy deposition $E$ in a module $i$ leads, with a probability $p=\epsilon_i(E)$ as derived in Section~\ref{sub:DetectionEfficiency}, to a positive trigger flag. Taking all geometrical effects as well as the modular trigger efficiency into account, a \muAcceptance\ of
\begin{linenomath}
\begin{align}
        \epsilon_{\rm{tot,MC veto-volume}} = (93.6 \pm 1.5)\%
\end{align}
\end{linenomath}
is obtained. Out of the 6.4\% of muons entering the veto volume undetected, more than a third of them (2.4\% of all) pass through geometrical gaps of the system. This reflects the geometrical coverage of the \muveto of 98\%. In addition, there are 0.9\% of all muons missed which pass through a module for which the data were not recorded during a small part of the total measurement time. This missing data is accounted for in the simulation by setting the trigger efficiency of this module to zero for the appropriate amount of simulated time. The statistical and systematic uncertainty of the \muAcceptance\ was evaluated assuming a set of different threshold values for the modular trigger efficiency. Simulating the complete muon flux as described in Section~\ref{sub:Geant4Setup}, it is noticeable that the number of triggered events is further increased by $\approx 25$\% due to muons passing outside the veto volume but having secondaries hitting at least one module.

Similarly to the \muAcceptance\ $\epsilon_{\rm{tot,MC veto-volume}}$, we also derived a \muAcceptance\ for a much smaller volume to be penetrated by muons, i.e. a sphere with a radius of 1\,m around the center of the cryostat was chosen.
This second case is intended as a more adapted efficiency value in consideration of the WIMP search. Whereas the first volume includes muons passing through some corners of the experimental setup with gaps due to mounting structures, the central sphere requires muon tracks that may produce secondary particles close to or within the cryostat. For these muons we obtain a \muAcceptance\ of
\begin{linenomath}
\begin{align}
        \epsilon_{\rm{tot,MC central sphere}} = (97.7 \pm 1.5)\%
\label{eq:effsyssphere}
\end{align}
\end{linenomath}
Since we rely on the same calibration and measured trigger efficiencies of the indiviual modules, the estimated error is of the same size as for $\epsilon_{\rm{tot,MC veto-volume}}$. The remaining inefficiency for this sample is almost entirely caused by the trigger inefficiencies of the individual modules and could thus be reduced by lowering the effective thresholds. As most of the \muind background in the bolometers is due to muons passing near the cryostat (see Section~\ref{sec:CoincTopo} and Fig.~\ref{fig:DistCryo}), $\epsilon_{\rm{tot,MC central sphere}}$ reflects better the veto efficiency for potential \muind background than $\epsilon_{\rm{tot,MC veto-volume}}$.

\subsection{Muon \muAcceptance\ derived from bolometer data}\label{sub:Eff-from-data}

\begin{figure}[tb]
 \begin{center}
 \includegraphics[width=0.48\textwidth]{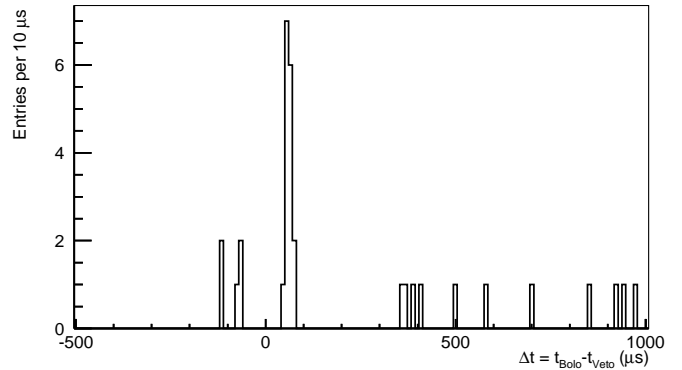}
\end{center}
 \caption{Time difference between bolometer muon candidates and their closest $\mu$-veto event. Note that two events with $\Delta t = -4.1$\,ms and $-12.1$\,ms are not shown.}
 \label{fig:Efficiency}
\end{figure}

The following approach avoids the uncertainties of the energy calibration and MC simulation and instead purely relies on data acquired with \EDW\ itself. It is fully independent of the methods described in Section~\ref{sub:resultMC} and leads to the most reliable results for the muon \muAcceptance\ . In this method, muon candidates are selected in the Ge bolometers and the efficiency is probed by a simple coincidence requirement. Above $6-7$\,MeV of energy deposition in the bolometers, the fraction of  events from the ambient radioactivity quickly drops and should be completely zero above 10\,MeV for anything besides throughgoing muons. To ensure suppression of all potential background in a single bolometer, an additional multiplicity requirement of at least two detectors was introduced. With a selection of $E_{\rm heat}>7$\,MeV and a multiplicity of $m_{\rm bolo}\geq 2$, a total of 34~$\mu$ candidate events was found in the bolometers. All of these events were detected in the \muveto with time
differences $\Delta t=t_{\rm bolo}-t_{\rm veto}$ as can be seen in Fig.~\ref{fig:Efficiency}. As will be discussed in detail in Section~\ref{sec:CoincInterval}, positive values of $\Delta t$ of the order of tens of $\mu$s are expected for perfect synchronization of the bolometer and \muvet . Such differences are due to the sampling rate of 100kS/s of the bolometer signals and the rise time of the ionization signal of the order of a few $\mu$s. Larger differences can occur for high energy events which saturate the ionization channel and make the time reconstruction more difficult. In contrast, there are 7 of the coincidences with an unphysical timing $\Delta t < 0$ where the fast scintillation signal of the $\mu$-veto was observed up to 12.1 ms after the signal in the bolometers (see Fig.~\ref{fig:Efficiency} and Section~\ref{sec:CoincInterval} for the discussion of the synchronization accuracy). Most of the events including the 7 events with unphysical timing, have reconstructed muon tracks (Section~\ref{mtrackreconst}) through the cryostat. A few of them show multiple hits indicating a shower in the whole system. Thus, with an expectation of 0.14 events due to accidental coincidences in the time interval $[-15,+1]$\,ms, all 34 events are considered as \muind coincidences.

The best estimate for the efficiency observing 34 out of 34 events is  $\epsilon_{\rm{tot,Data}}=100\%$. We determine a lower limit for the efficiency value by extracting the lowest efficiency value for which one expects to see 34/34 events in 10\% of the cases: $ P(\epsilon=x\, |\, 34/34\, \text{evts})=10 \% $. According to binomial statistics $P(k,n,p)=\binom{n}{k} p^{k}(1-p)^{n-k}$, where p corresponds to the efficiency $\epsilon$ and $n=k=34$, we transform this into
\begin{linenomath}
\begin{align}
\epsilon_{\rm{tot,Data}} \geqslant \sqrt[n]{0.1}=0.935 \label{eq:myEfficiency} \rm{\; at \; 90\%C.L.}
\end{align}
\end{linenomath}
Thus, this method yields a lower efficiency limit of 93.5\% entirely due to the statistical uncertainty. From the point of systematics, the candidate selection is biased to close muon tracks. However, this is also the case for the dominant fraction of all \muind events in the bolometers as will be discussed in Sec.\ref{sec:CoincTopo}. 
In conclusion, the method described here is supposed to be adequate for determining the rejection capability of the \muvet. The systematic uncertainty is estimated to be less than 1\% and the lower boundary of 93.5\% for the \muveto efficiency is regarded as a reliable albeit very weak boundary limited by the low statistics of the selected bolometer event sample. It is in full agreement with the \muAcceptance\ derived from MC as described in Section~\ref{sub:resultMC}. With the further data taking of \EDW, the statistics will increase and thus this method will also yield a more precise determination of $\epsilon_{\rm{tot,Data}}$.

For all following investigations of \muind background rates, we will use the muon \muAcceptance\ derived here, i.e. $\epsilon_{\rm{tot}}=\epsilon_{\rm{tot,Data}}$.

\subsection{Muon track reconstruction}
\label{mtrackreconst}
The information provided by the module number and time difference of the two PMT signals on each end of a plastic scintillator module are used to reconstruct the coordinates of the energy deposition in the $\mu$-veto system. Since there is no information about the position of the event with respect to the module width of 65\,cm and to the thickness of 5\,cm, the position perpendicular to the module axis is set to a random value between 0 and 65\,cm in order not to create discrete steps in geometrical distributions of the recorded events. The thickness of the module is neglected. For each event generating coordinate-signals in two modules, a track can be defined by connecting the two penetration points, including their relative timing. This track allows to reconstruct azimuth and zenith angles as well as the minimum distance of the muon track to the center of the cryostat.
\begin{figure}[tb]
 \centering
 \includegraphics[width = 0.48 \textwidth]{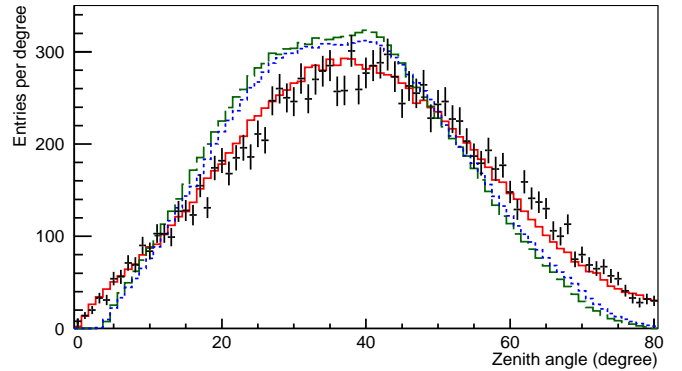}
 \caption{Zenith angle distribution of muon tracks in the EDELWEISS \muveto (black data points), in the \frejus proton decay detector (dotted blue) \cite{Rhode93} and of the corresponding simulated tracks (\EDW\ solid red; \frejus dashed green). All spectra are normalized to the \EDW\ data set. The simulation incorporates the full geometrical acceptance functions.}
 \label{fig:Zenith}
\end{figure}

The selection of events with exactly two energy depositions on different sides of the \muveto is supposed to be due to muons, since only those are penetrating enough to go through the whole experiment. The resulting angular dependent spectra are a negative imprint of the shielding properties of the mountain profile convoluted with the geometrical coverage and efficiencies of the experiment. In Fig.~\ref{fig:Zenith}, the measured zenith distribution of muons traversing the \EDW\ experiment is compared to MC simulations \cite{Ho07, Klu12} and previous measurements of the \frejus experiment \cite{Rhode93}. The MC simulation incorporates both the geometry and the single module efficiencies of the \EDW\ experiment as determined in section \ref{sub:DetectionEfficiency}. For better comparison of the angular distributions, all spectra are normalized to the \EDW\ data set. The MC simulation reproduces well the \EDW\ data (black data points). In contrast, the \frejus experiment \cite{Be89} has a different
geometry and hence the acceptance is significantly different
leading to a different zenith angle distribution as confirmed by simulated tracks with the according acceptance function. From Fig.~\ref{fig:Zenith}, one can see that the differences in the measurements of \EDW\ and \frejus are due only to the different acceptance functions. The simulated zenith angle distributions based on the same muon flux generator reproduce both angular distribution once the acceptance of the experiments is incorporated.

The actual profile of the \EDW\ data with a maximum at 35\degree\ is well understood with a large rock overburden of 1800\,m at zero degree and the smallest rock overburden at the flanks of the mountain. The spectral shape of the MC simulation is sensitive to the implemented energy thresholds of the individual modules. The consistency of the simulation with the \EDW\ data underlines that both the calibration and single module efficiency determination as well as the track reconstruction with an estimated reconstruction uncertainty of $\sigma_l=15$\,cm along the module axis are well controlled. The zenith spectrum has been cut at 80\degree\ since both the map of the mountain profile as well as the initial muon spectrum of the muon generator used in the MC have large uncertainties at such horizontal zenith angles.

\begin{figure}[tb]
 \centering
 \includegraphics[width=0.48 \textwidth]{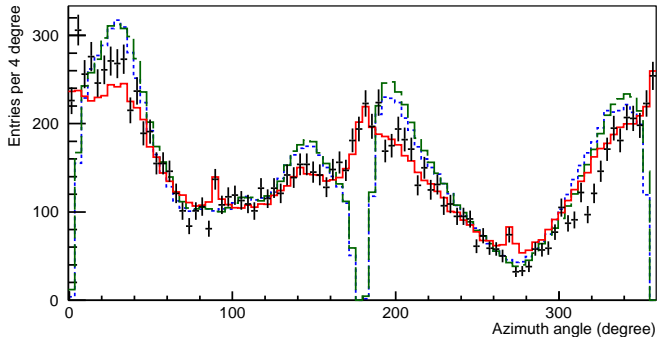}
 \caption{Azimuth angle distribution of muon tracks in the EDELWEISS \muveto (black data points), in the \frejus proton decay detector (dotted blue) \cite{Rhode93} and of simulated tracks through \EDW\ (solid red) and \frejus (dashed green). The \frejus detector has vanishing detection efficiency at 0 and 180 degrees. \frejus data and simulation results were normalized to the EDELWEISS data set.}
 \label{fig:Azimuth}
\end{figure}
The azimuth angle distribution (Fig.~\ref{fig:Azimuth}) also shows good agreement between the measured data in \EDW\ and the MC simulation modeling detected muon tracks. The zero degree direction is aligned with the laboratory direction. In the geographical coordinate system this corresponds to an offset of 16\degree\ to the north direction. There are two artificial peaks at 90\degree\ and 270\degree\ in the data. These peaks result from the fact that the modular position of events which have a TDC signal that would place them outside of a plastic scintillator module is set to be at the end of the module. For the top modules, the module ends are in plane with the North and South vertical scintillator modules giving reconstructed tracks with 90\degree\ or 270\degree\ azimuth angle.
The good agreement between the data and the simulation suggests that it describes well the spatial resolution and the proportion of tracks reconstructed near the module edges.

The profile measured by \frejus is rather similar to this work with a few exceptions. One immediately notes two dips in the angular spectrum. For azimuthal angles of 0\degree\ and 180\degree\ muons passed parallel to the detector planes of the \frejus experiment and could not be detected \cite{Be89}. Again, the same muon generator reproduces the two experimental azimuthal spectra once the according acceptance functions are incorporated into the simulations.

\begin{figure}[tb]
 \centering
 \includegraphics[width=0.48 \textwidth]{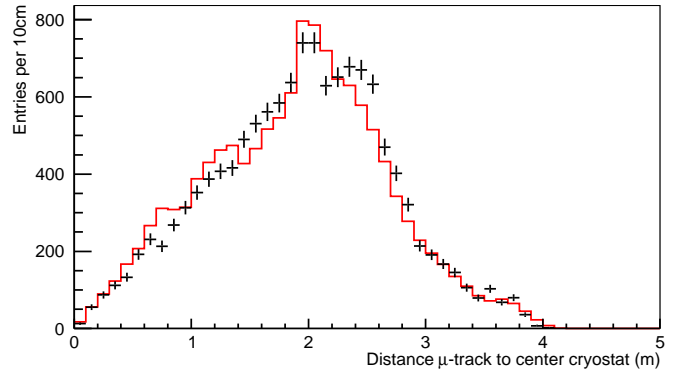}
 \caption{Muon track distance to the center of the cryostat. EDELWEISS $\mu$-candidates are shown as black data points and the MC simulation in solid red.}
 \label{fig:DistCryo}
\end{figure}
Another application of the $\mu$ track reconstruction is the identification of muons passing close to or penetrating the bolometers. The distribution of the minimum track distance to the center of the cryostat is shown in Fig.~\ref{fig:DistCryo}. The characteristic shape of the distribution originates from the number of muon tracks in a thin spherical shell with a certain fixed radius around the center of the cryostat and the fact whether this shell is contained within the \muveto or not. As the cryostat is 1.9\,m away from the top and the four upper sidewalls, this distance corresponds to the maximum of the muon track distribution. Afterwards the number of muon tracks within the spherical shell still increases, but the number of detected traces decreases again. The maximal distance of a muon veto module to the center of the cryostat is $d\approx 4$\,m. The simulation reproduces very well the data and underlines the reliability of the muon track reconstruction.

\subsection{Determination of the muon flux}
\label{sec:muonFlux}
To conclude, we determine the integral muon flux through a horizontal surface. Requiring a module being hit in at least two surfaces of the \muvet, we accumulated 18503 muon candidates which corresponds to a measured rate of $\Gamma_{\mu-\text{cand.}} = (108.7\pm 0.8)/d$. Due to the rather cubic geometry and the individual detection efficiencies of each module, the overall detection acceptance for a given muon selection must be determined through simulations. For the applied muon selection cuts, a combined detection acceptance $a_{\text{MC}}=20.0\pm 0.4 \, {\rm candidates}/(\mu/m^2)$ was derived from simulations accounting for the geometry of the setup and modular efficiencies $\epsilon^{\mu}_i$. This leads to an acceptance-corrected total flux of \muind events of
\begin{linenomath}
\begin{align}
 \Phi_{\mu}&=\Gamma_{\mu-\text{cand.}}/a_{\text{MC}} \nonumber \\ &= 5.4\pm 0.2(\text{stat.})^{+0.5}_{-0.9}(\text{syst.})\, \mu/m^2/d.
 \label{eq:muflux}
\end{align}
\end{linenomath}
The systematic uncertainty is made up of three contributions. The first one arises from background events not induced by muons but being in coincidence in modules of two geometrical surfaces. This may happen in corners of the \muvet . To check their influence, we analyzed subsamples where adjacent surfaces are not allowed. This reduces $\Gamma_{\mu-\text{cand.}}$ as well as $a_{\text{MC}}$ leading to an overall reduction of $\Phi_{\mu}$ by 10\,\%. As described in Section~\ref{sub:DetectionEfficiency}, the modular efficiencies $\epsilon^{\mu}_i$ depend on the calibration parameters. Allowing a variation of $\pm20$\,\% leads to another 10\,\% uncertainty. Third, there is a remaining uncertainty of the derived flux due to modeling the muon flux based on single muons produced on the hemisphere and neglecting muon bundles which were measured in the \frejus experiment contributing 5\% to the overall muon flux \cite{Be89}. In our simulations, the average number of muons per event is only slightly larger than
one due to secondary muons produced in the rock between the start hemisphere and the lab. Deriving alternatively from the 18503 measured muon candidates the muon flux through a sphere leads to $a_{\text{MC}}^{\text{sph}}=16.4\pm 0.4$ candidates/$(\mu/m^2)$ and subsequently $\Phi_{\mu}^{\text{sph}}= 6.6\, \mu/m^2/d$.

The measured flux as given in equ.~\ref{eq:muflux} is slightly higher but yet compatible with the measurements of the \frejus experiment. To compare the values, the different definitions and cuts have to be taken into account. The event rate determined here corresponds to the sum of events with muon multiplicity $m \geq 1$, including muon bundles. In addition, no zenith angle cut of $\Theta < 60 \degree$ as in \cite{Be89} was applied in our measurement. Since it is not stated in \cite{Be89} whether the muon flux was measured for a horizontal area or a sphere, we give the appropriate values for both assumptions here. Interpreting the quoted flux in \cite{Be89} as a flux through a horizontal plane and correcting it for the inefficiency of the $60 \degree$ cut, evaluated from our measurements, a value of $\Phi_{\text{Fr\'{e}jus}}^{\rm horizontal}=(5.2\pm 0.1_{\text{stat.}})\mu/m^2/d$ is calculated. The alternative interpretation of the values in \cite{Be89} as the flux through a sphere would lead
to a smaller flux of $\Phi_{\text{Fr\'{e}jus}}^{\rm horizontal} =(4.2\pm 0.1_{\text{stat.}})\mu/m^2/d$ to be compared with the value given in equ.~\ref{eq:muflux}.

\section{Coincidences in the \muveto and bolometers}
\label{sec:Coincs}

For WIMP searches, every background component has to be well understood and maximally suppressed. In this section, the muon-induced event population in the bolometers is studied in situ, for the specific geometrical arrangement of bolometers during the data acquisition period from March 2009 to May 2010. In addition to quantifying the muon-induced neutron component for the WIMP analysis \cite{Arm11}, the topology of muon-induced events is investigated. With this derived information, an estimate of muon-induced neutrons can be deduced for the next phase of the experiment, \EDDD. The investigation is performed using a newly developed data framework, KData \cite{Cox12} that allows the event-based storage and analysis of both bolometer and $\mu$-veto data in the same structure.

\subsection{Determination of the coincidence interval}
\label{sec:CoincInterval}
Searching for delayed coincidences between a prompt muon veto hit and a secondary bolometer event, a typical time delay $\Delta t=t_{\rm bolo}-t_{\rm veto}$ of a few tens of microseconds is expected.
\begin{figure*}[tb!]
 \centering
 \includegraphics[width=0.7 \textwidth]{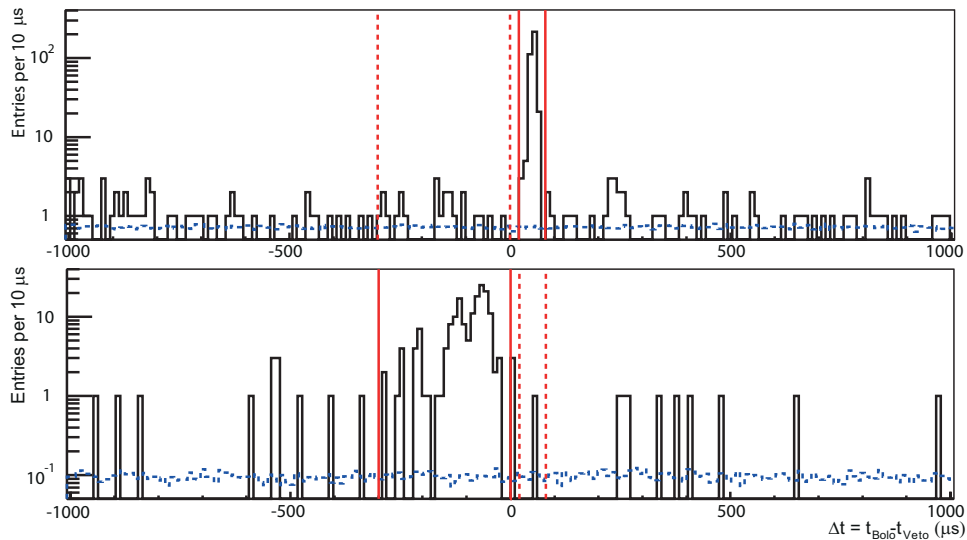}
 \caption{Time difference between bolometer events and hits in the \muvet . (a) data taken after September 2009, (b) data taken from March to September 2009. Black indicates coincidences while the blue dotted line corresponds to the expected rate of accidental background. The vertical lines represent timing cuts applied for further investigation of coincidences. Solid lines indicate the applied cut for the appropriate data period while the dotted lines describe the cut used for the other data period.}
 \label{fig:deltat}
\end{figure*}
This time difference is mainly due to the rise time of the bolometer pulses and their sampling rate of 100\,kS/s. The accuracy of the synchronization of the separate acquisition systems leads to a broadening of a few tens of microseconds. The overall recorded time difference as shown in Fig.~\ref{fig:deltat} is thus not caused by physics processes such as thermalization of neutrons. The exact experimental coincidence window to select \muind signals is determined by the $\Delta t$ distribution itself. In the following, we select events where at least two modules were hit ($m_{\rm veto}\geq 2$ with at least one ADC or TDC entry per module being non-zero) leading to an event rate of 0.12\,Hz. This is a much stricter requirement than in vetoing potential backgrounds in the \dm search where bolometer events are rejected whenever there is \textit{any} event recorded in the \muveto within a $\pm 1$\,ms time interval. With an  event rate of 0.41\,Hz in the \muvet , the
applied dead time amounts to less than 0.1\,\% of the total measuring time. Having a muon candidate with $m_{\rm veto}\geq 2$, we then extract the time difference $\Delta t$ to the nearest event in the bolometer data set.

This time difference is shown in Fig.~\ref{fig:deltat} for two distinct periods of data taking. Fig.~\ref{fig:deltat}a shows the distribution for the data after September 2009 with fully functioning synchronization between \muveto and bolometers. A clear peak of \muind coincidences can be identified and a time interval of 20\,$\mu s \leq \Delta t \leq 80$\,$\mu s$ was chosen maximizing the signal-to-background ratio containing at least $97.5$\,\% of the peak population. For the early data set from March 2009 to mid-September 2009 corresponding to about 1/3 of the total exposure, some data losses in the optical transfer of the 48-bit clock from bolometers to the \muveto degraded the synchronization performance. An offline correction to the synchronization could reproduce a peak in $\Delta t$, but at slightly unphysical time differences (Fig.~\ref{fig:deltat}b). The corresponding interval for coincidences was derived to be -300$\,\mu s \leq \Delta t \leq 0$\,$\mu s$. However, not all of the synchronization
errors could be corrected in the offline analysis. A detailed investigation of coincidences with synchronization problems resulted in the quantification of $(12\pm 8)\,\%$ of all cases with $\Delta t<-1$\,ms (including the two outliers at $\Delta t=-12.1$\,ms and $\Delta t=-4.1$\,ms in Section~\ref{sub:Eff-from-data}). An attributed uncertainty of a few ms in synchronizing the systems is much larger than the vetoing interval of $\pm1$\,ms around each veto hit. When deriving the \muind background in the WIMP search in Section~\ref{sec:CoincBg}, we therefore conservatively treat these periods with an exposure of $(17\pm 11)$\kgd as bolometer data without functioning \muvet .

With the two coincidence intervals as shown in Fig.~\ref{fig:deltat}, a total of  $76 + 158 = 234$ events summing over the 1st and 2nd period were selected for further investigation. An accidental background of $6.75\pm 0.03$ events was determined in a two second wide window with an offset of 5 seconds before the veto event. This corresponds to an achieved signal to noise ratio of better than 33/1. After subtraction of accidental background, the remaining excess of coincidences in \EDD\ is entirely due to events induced by muons.

\subsection{Topology of \muind bolometer events}\label{sec:CoincTopo}

\begin{figure}[tb]
 \centering
 \includegraphics[width=0.48 \textwidth]{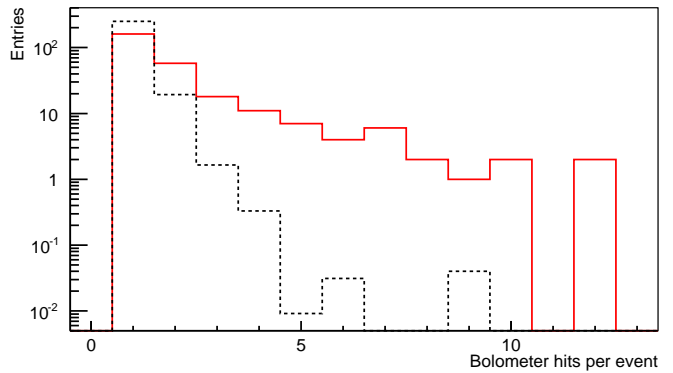}
 \caption{Bolometer multiplicity for \muind events (solid red). For comparison, a scaled histogram of the multiplicity of
all bolometer events is superimposed (dotted black).}
 \label{fig:Multi}
\end{figure}

A WIMP scattering off a Ge nucleus would lead to a single scatter event in a bolometer, i.e. a bolometer multiplicity $m_{\rm bolo}=1$. Any event with $m_{\rm bolo}>1$ can therefore be rejected. The multiplicity of bolometer hits in a bolometric array is thus a potentially powerful tool to reject background, depending on the nature of the background as well as on the total detector mass and the granularity of the crystal array. The multiplicity of the 234 extracted $\mu$-induced bolometer events is shown in Fig.~\ref{fig:Multi}. For the specific setup of \EDD, $(46.6\pm0.6)$\,\% of all muon-induced events in the complete energy range would be rejected in a WIMP search due to $m_{{\rm bolo,}\mu}>1$, with an average bolometer multiplicity of $\langle m_{{\rm bolo,}\mu}\rangle = 2.27 \pm 0.18$. This is in clear contrast to $\langle m_{\rm  bolo,all}\rangle =1.05$ for all bolometer events, with the $m_{\rm  bolo,all}$ distribution also shown in Fig.~\ref{fig:Multi}.

The fraction of events with $m_{{\rm bolo,}\mu}>1$ strongly depends on the geometrical setup. This can be seen in a comparison of the present results with the simulations based on a fully equipped cryostat with 120 Ge crystals, each with a mass of 320\,g \cite{Ho07}, adding up to 38.4\,kg. As described in Section~\ref{sec:BolometerSystem}, the current study incorporates only 10 ID detectors with a total mass of 3.6\,kg. In the ideally packed array used in the simulations, a fraction of 84\,\% of all \muind events have a multiplicity of $m_{{\rm bolo,}\mu}>1$. Thus, with increasing numbers of bolometers in a densely packed array, a \muind background can be efficiently suppressed due to its high multiplicity.
\begin{figure}[tb]
 \centering
	\includegraphics[width=0.48 \textwidth]{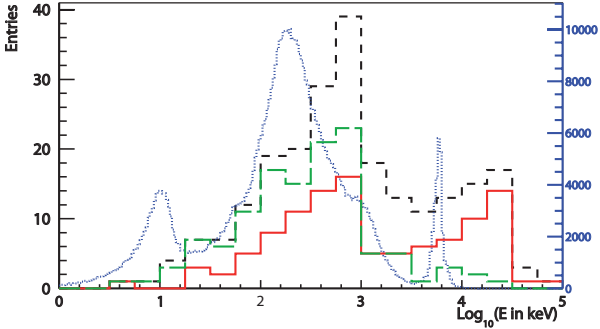}
 \caption{Sum of energy deposition in the bolometers for the 234 coincidence events (dashed black - 4 bins per decade) as well as for \muind single bolometer hits (long-dashed green) and background events in individual bolometers (dotted blue - 64 bins per decade). For comparison, a subsample of 109 events is shown for which a muon track could be reconstructed (solid red, see text for more details). }
 \label{fig:ESPec}
\end{figure}

Furthermore, \muind background events exhibit an energy spectrum which is very different to an exponentially decreasing recoil spectrum expected from WIMP scattering. Fig.~\ref{fig:ESPec} shows the sum energy of all energy depositions in the detectors associated with a \muind event. The spectrum is composed of two main regions. Up to 1\,MeV the energy deposition in the bolometers is caused by single or multiple energy depositions of secondaries, mainly electrons and gammas. From 50\,keV to several 100\, keV the dominant interaction is Compton scattering. The second feature is a broad peak between 5 and 100\,MeV with a maximum around 20\,MeV caused by muons passing through the Ge crystals (energy loss of muons in Ge $\approx$\,10\,MeV/cm). For those events, the energy deposition is mainly dependent on the path length of the muon in the crystals. For single scatter events (Fig.~\ref{fig:ESPec} long-dashed green) this feature is strongly suppressed, indicating that these are mainly caused by secondaries.
Last, there is a small fraction of events below 50\,keV, where the dominant component is elastic scattering of \muind neutrons as will be discussed later. Fig.~\ref{fig:ESPec} also shows the energy deposition in the bolometers for background events (dotted blue histogram) which clearly contrasts the \muind distribution. The peaks in this spectrum can be associated to $\alpha$ decays of $^{210}$Po (5.4\,MeV), which can be further suppressed by a fiducial volume cut, a Compton backscattering bump above 100\,keV as well as the 10\,keV gamma line from cosmogenic activation.

\begin{figure}[tb]
 \centering
 \includegraphics[width=0.48 \textwidth]{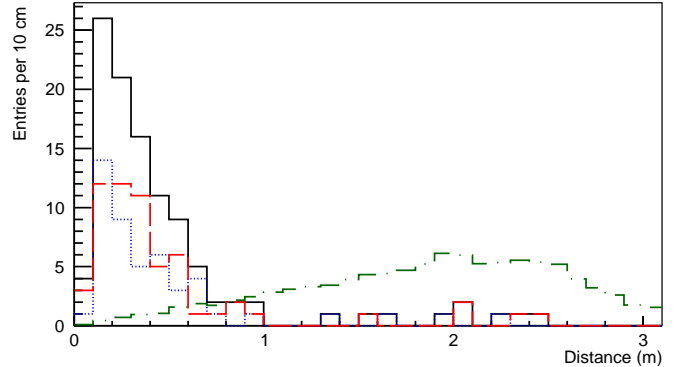}
 \caption{Distance to the center of the cryostat for 109 coincidence events with track reconstruction (solid black), with energy depositions in the bolometers below (dotted blue) and above 1\,MeV (dashed red). Also shown is the scaled distribution of muon events identified in the \muveto without requiring a coincidence (dash-dotted green) with its maximum at $\approx 2$\,m as described in Section~\ref{mtrackreconst} and Fig.~\ref{fig:DistCryo}.}
 \label{fig:DSPec}
\end{figure}
Apart from the bolometer multiplicity and energy deposition of \muind events, the distance of the muon track from the center of the cryostat could be deduced from the information in the muon veto modules. Fig.~\ref{fig:DSPec} shows the distance of 109 out of the 234 sequential events. For these 109 events, the muon track could be reconstructed unambiguously due to full spatial information in exactly two of the veto modules. More than 90\% of these \muind bolometer events are produced by muons passing within a distance of less than 1\,m to the cryostat's center whereas the overall distribution of all identified muons is much broader with a maximum around $d=2$\,m. Although \muind bolometer events have short track distances to the cryostat, it is important to note that this does not translate directly into muons hitting the Ge crystals with large energy depositions. As can be seen in Fig.~\ref{fig:DSPec} the distances for events with energy depositions in the bolometers below (dotted blue) and above 1\,
MeV (
dashed red) have almost identical distributions. Thus, muons even passing through the cryostat lead to energy depositions below 1\,MeV, reaching the typical WIMP range of tens of keV, indicating energy deposition by secondary particles. On the other hand, distances of $d>0.7$\,m correspond to muons passing outside the cryostat, where secondary electrons and gammas are unlikely to reach the Ge bolometers due to the various shields. In this case, coincident events can be interpreted as \muind neutrons scattering off the Ge nuclei.

\begin{figure}[tb]
 \centering
 \includegraphics[width=0.48 \textwidth]{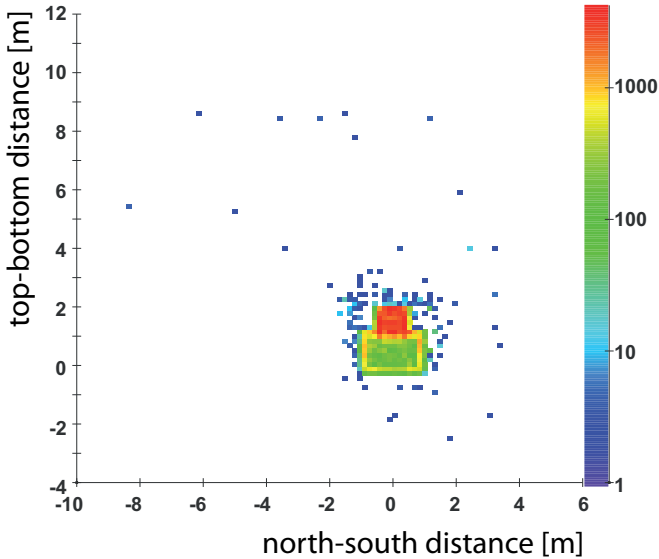}
 \caption{Simulation of production vertices of neutrons in \muind showers leading to an energy deposition by nuclear recoils in the bolometers \cite{Ho07}. Shown is a side projection of the underground laboratory LSM. Most of the neutrons are produced in the Pb shield and the steel of the support structure.}
 \label{fig:Horn}
\end{figure}
The requirement for reconstructing muon tracks may introduce a slight bias in selecting \muind events with high energy deposition in the bolometers as can be seen in Fig.~\ref{fig:ESPec}. Low energy depositions originating from \muind neutrons might also be connected to muons passing outside the \muvet . However, the production vertices for these neutrons are dominantly positioned in the Pb shielding as can be seen in Fig.~\ref{fig:Horn}. From these simulations, one expects only 0.05\,\% of all neutrons produced outside of the \muvet , while more than 90\,\% are produced within the lead shield or the cryostat itself. In conclusion, \muind bolometer events (including potential neutron scattering events) are mainly caused by muons with a track of less than 1\,m distance from the center of the cryostat and thus potentially can be vetoed with the \muvet .

\subsection{\Muind background for \dm search}\label{sec:CoincBg}

\begin{figure}[tb]
 \centering
 \includegraphics[width=0.43 \textwidth, trim = 5mm 0mm 15mm 0mm]{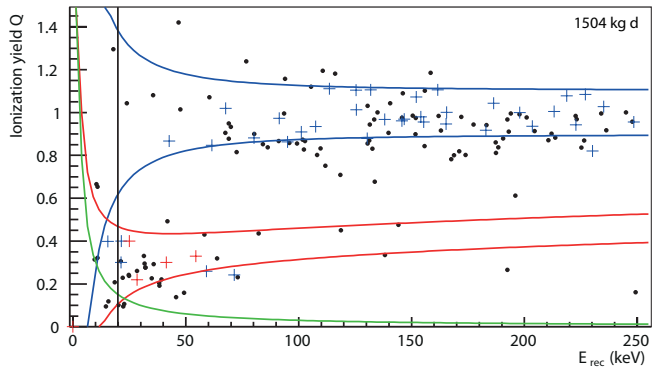}
 \caption{Scatter plot of bolometer hits in the coincidence analysis for energies below 250\,keV. Fiducial events that are expected to have a full charge collection and thus a reliable value for Q are drawn as crosses (blue and red). The events marked in red satisfy all WIMP criteria. Other events are shown as black dots. The lines correspond to the 90\,\% C.L. gamma region (blue), the 90\,\% C.L. nuclear recoil band (red), the analysis threshold of 20\,keV used in the WIMP analysis (black) and an ionization threshold for fiducial events of 3\,keV (green). Except for the analysis threshold, all resolutions and bands are time and bolometer dependent. For illustration, the time-averaged bands of an exemplary detector, ID401, are drawn here.}
 \label{fig:ScatterCoincEvents}
\end{figure}

The \muind background for the WIMP analysis is determined as the product of the rate of \muind single hit events in the nuclear recoil band, the accumulated exposure of 384\,\kgd and the \muveto inefficiency ($1-\epsilon_{\rm{tot}}$) as determined in eq.~(\ref{eq:myEfficiency}). For a DM search, it is crucial to have the highest muon detection efficiency possible. Hence, in this analysis we abandon the requirement for muon candidates as used in Sections~\ref{sec:CoincInterval} and \ref{sec:CoincTopo}, i.e. having hits in more than one veto module. The total number of coincidences then moderately increased from $N_C^{m>1} = 234$ to $N_C = 283$ events, while the rate of accidental background increased by a factor of 3.5 from $N_A^{m>1} = 6.75\pm 0.03$ to $N_A = 23.61\pm 0.06$. As for the 234 coincidences studied before, there was no rejection of periods with high baseline noise nor any selection of fiducial bolometer events. Thus, the total exposure is much larger than in the WIMP analysis
and amounts to $1504 \, \rm{kg \cdot d}$. With a net number of $N = N_C - N_A = 259\pm 17$ coincidences, the rate of muon-induced events in the complete energy range, not discriminating between electron and nuclear recoils, amounts to
\begin{linenomath}
\begin{align}
 \Gamma^{\mu\rm{-ind.~evts}}&=(0.172 \pm 0.001)\, \rm{evts}/(\rm{kg \cdot d}).
\end{align}
\end{linenomath}
The rate of \muind background events within \EDD\ is in good agreement with earlier investigations \cite{AC_diss} based on a much smaller event sample and slightly different selection cuts. Fig.~\ref{fig:ScatterCoincEvents} shows the events as individual bolometer hits with an energy deposition below 250\,keV. Applying a fiducial volume cut inferring reliable charge collection and thus ionization yield and a $\chi^2$-cut comparing heat and ionization pulse fits with their template pulses, bolometer hits shown as dots are rejected. After a time-averaged 90\,\% C.L. nuclear recoil band selection applied for each single detector, seven bolometer hits remain (see Fig.~\ref{fig:ScatterCoincEvents} within the red bands). Out of these seven, six are above the 20\,keV analysis threshold, and only four (shown as red crosses) are single bolometer hits without a coincident hit in another bolometer as required for a WIMP candidate. (The bolometer hit at 60\,keV is in coincidence with another one, the hit at 70\,keV is
outside the predefined nuclear recoil band.)  Three of these events have reconstructed muon tracks going through the lead shield and even through the cryostat itself. The fourth event is in coincidence with an event in the \muvet, where 11 plastic scintillator modules have been hit, indicating a \muind shower.  Fig.~\ref{fig:ScatterCoincEvents} shows individual bolometer hits. Rates in Table \ref{tab:ResultsRateSingleMuon} refer to bolometer events with potentially more than one bolometer hit.

Applying all cuts required for WIMP candidates results in an effective fiducial exposure of 481\,kg$\cdot$d. Thus, the rate of \muind WIMP-like events amounts to
\begin{linenomath}
\begin{align}
 \Gamma^{\mu-n}&=(0.008 ^{+0.005}_{-0.004})\, \rm{evts}/(\rm{kg \cdot d}). \label{eq:WIMPlikeRate}
\end{align}
\end{linenomath}
Assuming a conservative veto efficiency as in Eq.~\ref{eq:myEfficiency}, the rate of muon-induced WIMP-like events amounts to an irreducible background rate of $\Gamma^{\mu-n}_{irred} < 5.8\cdot 10^{-4} \, \rm{evts}/(\rm{kg \cdot d})$ at 90\,\%\,C.L.
\begin{table}

\centering
\caption{Rates of coincidences. All results are given after background subtraction. The cuts on recoil energy $\rm E_{rec}^{hit}$ and Q-value $Q_{rec}^{hit}$ are fulfilled, when at least one bolometer hit within an event fulfills the condition. For fiducial nuclear recoil (NR) events, the 90\,\% signal region is defined following the Lindhard parametrization~\cite{Lin63}.}

 
{
\fontfamily{iwona}\fontseries{c}\fontsize{7}{8}\selectfont

\begin{tabular}{lrrr}\toprule

Selection cut & $\Gamma_{\mu\rm{-ind}}$ $(\rm{evts}/\rm{kg / d})$& Exposure $(\rm{kg\, d})$  \\ \midrule
None  & $0.172 \pm 0.012$ & $1504$\\
$\rm E_{rec}^{hit} < 250$\,keV  & $ 0.095\pm0.009 $ & $1504$\\
$\rm E_{rec}^{hit} < 250$\,keV; $\rm Q_{rec}^{hit} < 0.6$ & $ 0.035\pm 0.005$ & $1504$\\
$\rm E_{rec}^{hit} < 250$\,keV; fid. NR   & $ 0.017 \pm 0.006 $& $481$ \\
$\rm E_{rec}^{hit} < 250$\,keV; fid. NR; single   & $0.008 ^{+0.005}_{-0.004}$ &$481 $\\
\label{tab:ResultsRateSingleMuon}
\end{tabular}
}
\end{table}

\small

From Equations \ref{eq:myEfficiency} and \ref{eq:WIMPlikeRate} for the $\mu$-veto detection efficiency and the rate of \muind WIMP-like events, the number of expected unvetoed \muind single scatter neutron events $N^{\mu-n}$ within the WIMP analysis is calculated according to
\begin{linenomath}
\begin{align}
 N^{\mu-n}&=M_{\rm{Exp}}^{\rm{B+V}} \Gamma^{\mu-n} (1-\epsilon_{\rm{tot}})+ M_{\rm{Exp}}^{\rm{B}} \Gamma^{\mu-n} = 0.40 
 \label{eq:edw2expect}
\end{align}
\end{linenomath}
Here $M_{\rm{Exp}}^{\rm{B}}=38\pm 11$\,\kgd denotes the sum of exposure of a short period where only bolometer data was acquired ($21$\,kg$\cdot$d) and the earlier period where a hardware malfunction led to a faulty synchronization in 12\,\% of the measuring time ($17\pm 11$\,kg$\cdot$d). $M_{\rm{Exp}}^{\rm{B+V}}=384\,\rm{kg \cdot d} - M_{\rm{Exp}}^{\rm{B}}$ is the exposure of the WIMP search data set, when both acquisition systems (bolometer and \muvet) were running with full synchronization. The above expectation value $N^{\mu-n}$ is dominated by the contribution from $M_{\rm{Exp}}^{\rm{B}}$. Taking into account all uncertainties on $\Gamma^{\mu-n}$, $\epsilon_{\rm{tot}}$ and $M_{\rm{Exp}}^{\rm{B}}$ we derive an upper limit of $N^{\mu-n} < 0.72$~events (90\,\%\,C.L.). Comparing this result with other background estimates for the \EDD\ WIMP search from misidentified gamma events $N^{\gamma}<0.9$, from surface events $N^{\beta}<0.3$ \cite{Arm11} and from ambient neutrons $N^n<3.1$ \cite{EDWbg12}, \muind background contributes less than 20\,\% to the total background. Note that this is dominantly due to including periods in the WIMP search where the \muveto was not synchronized properly to the bolometer system. For future searches as in \EDDD , such periods will not be included and thus an even better rejection of \muind background can be achieved.

\section{Expected background for \EDDD\ and EURECA}
\label{sec:EURECA}

For the forthcoming phase of \EDDD\ as well as future experiments, an accurate number for the \muind neutron background can be predicted only with full modeling of the specific geometry due to the suppression of multiple scattering signatures in a densely packed detector array and the benefits from additional shielding. A larger number of detectors, such as the forty 800\,g detectors planned for \EDDD, will increase the granularity and hence the number of multiple bolometer hits. Also, the effective area of Ge bolometers exposed to the muon flux remains approximately constant while the detector mass increases proportionally with the number of detector planes. Furthermore, the introduction of an additional PE shield inside the cryostat is going to moderate the neutron flux induced in the lead. All effects will reduce the rate of muon-induced WIMP-like background events.

Taking the measured rate of \muind nuclear recoil events (Eq.~\ref{eq:WIMPlikeRate}) and the derived efficiency of the \muveto of 97.7\,\% as in Eq.~\ref{eq:effsyssphere}, a total of $N^{\mu-n}_{3000 kg\cdot d}=0.6^{+0.7}_{-0.6}$ (90\,\%\,C.L.) irreducible background events due to \muind neutrons are projected for \EDDD\ for a 6-month exposure of 3000$\, \rm{kg \cdot d}$. Both the background rate and the muon detection efficiency will be measured in the ongoing analysis of the \EDDD\ experiment and in parallel modeled with its specific new setup and better shielding.

Going further to ton-scale experiments such as the envisaged EURECA experiment, the \muind neutron background could be of sizable contribution. The strategy of reducing it with respect to the \EDW\ design is to reduce the high-Z material close to the bolometers as well as including a highly efficient water \v{C}erenkov active shield around the cryostat \cite{Eureca_Kraus_idm10}. 
\section{Conclusion and outlook}

The angular dependence of the muon flux through the \EDW\ experiment has been analyzed and is well modeled by MC simulations. Two independent methods to determine the \muveto efficiency have been presented, giving results limited so far by small statistics. A conservative value of $\epsilon_{\rm{tot}} > 93.5\,\%$ has been used for the calculation of \muind background in the WIMP search data set. The data of the bolometer array together with the modular structure of the \muveto allowed a detailed study of the topology of \muind events within \EDW. The analysis of \muind background has been performed in the specific setup of the data acquisition of 2009/2010 with 10 Ge bolometers. For this configuration, the rate of muon-induced events that mimic WIMPs has been determined to $\Gamma^{\mu-n}=(0.008 ^{+0.005}_{-0.004})$ events/(kg\,$\cdot$\,d). In conclusion, the \muind background with an expectation of $N^{\mu-n}< 0.72$ events within this WIMP search was small compared to other backgrounds. Thus,
only a minor fraction of the five events observed in the \dm search \cite{Arm11} is assumed to be due to \muind interactions.

The feasibility of a background free data taking of at least ${3000 \kgdmath}$ (\EDDD\ 6 months) has been shown with a conservative extrapolation from current data. Expected benefits from larger bolometer granularity and hence larger multiplicity of events as well as from the added shielding have been briefly discussed and should enable background free operation beyond a sixth month exposure. However, future ton-scale experiments, like the EURECA experiment, need different shielding concepts with less high-Z material close to the bolometers.

\section{Acknowledgements}
\label{sec:Acknowledge}
The help of the technical staff of the Laboratoire Souterrain de Modane and the participant laboratories is gratefully acknowledged. This project is funded in part by the German ministry of science and education (BMBF) within the "Verbundforschung Astroteilchenphysik" grant 05A11VK2,  by the Helmholtz Alliance for Astroparticle Physics (HAP) funded by the Initiative and Networking Fund of the Helmholtz Assocication, by the French Agence Nationale de la Recherche under the contracts ANR-06-BLAN-0376-01 and ANR-10-BLAN-0422-03 and by the Science and Technology Facilities Council (UK).
The dedicated neutron measurements (ambient and muon-induced) were partly supported by the EU contract RII3-CT-2004-506222 and the Russian Foundation for Basic Research.

\bibliographystyle{plain}
\bibliography{muonBackgroundStudyEtAlTitle}

\end{document}